\documentclass[aps,twocolumn,amsmath,amssymb,showpacs,prb,superscriptaddress,unsortedaddress]{revtex4}

\usepackage{epsf}
\usepackage{graphicx}
\usepackage{sidecap}

\begin{document}

\title{Topological surface states and Dirac point tuning in ternary topological insulators}

\author{M.~Neupane}\affiliation {Joseph Henry Laboratory, Department of Physics, Princeton University, Princeton, New Jersey 08544, USA}

\author{S.-Y.~Xu}\affiliation {Joseph Henry Laboratory, Department of Physics, Princeton University, Princeton, New Jersey 08544, USA}

\author{L.~A.~Wray}\affiliation {Joseph Henry Laboratory, Department of Physics, Princeton University, Princeton, New Jersey 08544, USA}\affiliation {Advanced Light Source, Lawrence Berkeley National Laboratory, Berkeley, California 94305, USA}

\author{A.~Petersen}\affiliation {Joseph Henry Laboratory, Department of Physics, Princeton University, Princeton, New Jersey 08544, USA}

\author{R.~Shankar}\affiliation {Joseph Henry Laboratory, Department of Physics, Princeton University, Princeton, New Jersey 08544, USA}

\author{N.~Alidoust}\affiliation {Joseph Henry Laboratory, Department of Physics, Princeton University, Princeton, New Jersey 08544, USA}

\author{Chang~Liu}\affiliation {Joseph Henry Laboratory, Department of Physics, Princeton University, Princeton, New Jersey 08544, USA}

\author{A.~Fedorov}\affiliation {Advanced Light Source, Lawrence Berkeley National Laboratory, Berkeley, California 94305, USA}

\author{H.~Ji}\affiliation {Department of Chemistry, Princeton University, Princeton, New Jersey 08544, USA}

\author{J.~M.~Allred}\affiliation {Department of Chemistry, Princeton University, Princeton, New Jersey 08544, USA}

\author{Y.~S.~Hor}\affiliation {Department of Chemistry, Princeton University, Princeton, New Jersey 08544, USA}

\author{T.-R. Chang}
\affiliation{Department of Physics, National Tsing Hua University, Hsinchu 30013, Taiwan}

\author{H.-T. Jeng}
\affiliation{Department of Physics, National Tsing Hua University, Hsinchu 30013, Taiwan}
\affiliation{Institute of Physics, Academia Sinica, Taipei 11529, Taiwan}

\author{H.~Lin}\affiliation {Department of Physics, Northeastern University, Boston, Massachusetts 02115, USA}

\author{A.~Bansil}\affiliation {Department of Physics, Northeastern University, Boston, Massachusetts 02115, USA}

\author{R.~J.~Cava}\affiliation {Department of Chemistry, Princeton University, Princeton, New Jersey 08544, USA}

\author{M.~Z.~Hasan}\affiliation {Joseph Henry Laboratory, Department of Physics, Princeton University, Princeton, New Jersey 08544, USA}

\begin{abstract}

Using angle-resolved photoemission spectroscopy, we report electronic structure for representative members of ternary topological insulators. We show that several members of this family, such as Bi$_2$Se$_2$Te, Bi$_2$Te$_2$Se, and GeBi$_2$Te$_4$, exhibit a singly degenerate Dirac-like surface state, while Bi$_2$Se$_2$S is a fully gapped insulator with no measurable surface state. One of these compounds, Bi$_2$Se$_2$Te, shows tunable surface state dispersion upon its electronic alloying with Sb (Sb$_x$Bi$_{2-x}$Se$_2$Te series). Other members of the ternary family such as GeBi$_2$Te$_4$ and BiTe$_{1.5}$S$_{1.5}$ show an in-gap surface Dirac point, the former of which has been predicted to show nonzero weak topological invariants such as (1;111); thus belonging to a different topological class than BiTe$_{1.5}$S$_{1.5}$. The measured band structure presented here will be a valuable guide for interpreting transport, thermoelectric, and thermopower measurements on these compounds. The unique surface band topology observed in these compounds contributes towards identifying designer materials with desired flexibility needed for thermoelectric and spintronic device fabrication.

\end{abstract}
\date{\today}
\maketitle

\section{Introduction}

A topological insulator (TI), as experimentally realized in bismuth-based materials, is a novel electronic state of quantum matter characterized by a bulk-insulating band gap and spin-polarized metallic surface states. \cite{Kane PRL, David Nature08, Hasan, SCZhang, Suyang_1, Ran Nature physics, David Science BiSb, Matthew Nature physics BiSe, Chen Science BiTe, David Nature tunable, Pedram Nature BiSb, Hor PRB BiSe, Essin PRL Magnetic, Galvanic effect, Yu Science QAH, Qi Science Monopole, Linder PRL Superconductivity, Liang Fu PRL Superconductivity, Phuan, Hor arXiv BiTe superconducting} Owing to time reversal symmetry, topological surface states are protected from backscattering and localization in the presence of weak perturbation, resulting in spin currents with reduced dissipation. On the other hand, bismuth-based materials are also being studied for enhanced thermoelectric device performance. \cite{Moore} Therefore, it is of general importance to study the band structure of these materials as a starting point. Using angle-resolved photoemission spectroscopy (ARPES) and spin-resolved ARPES, several Bi-based topological insulators have been identified, such as the Bi$_{1-x}$Sb$_x$ alloys \cite{David Nature08, David Science BiSb}, the Bi$_2X_3$ ($X$ = Se, Te) series and their derivatives.\cite{Matthew Nature physics BiSe, Chen Science BiTe} Although significant efforts have been made to realize multifunctional electronic properties in the existing materials, little success has been obtained so far due to residual bulk conduction.\cite{Phuan, Hor arXiv BiTe superconducting, Suyang} This led to the search for other topological materials, which might potentially be optimized for the realization of functional devices.

Recently, ternary topological insulators such as Bi$_2$Se$_2$Te, Bi$_2$Te$_2$Se, Bi$_2$Te$_2$S, GeBi$_2$Te$_4$, and PbBi$_4$Te$_{7}$ have been theoretically predicted to feature multifunctional and flexible electronic structures. \cite{Suyang_1, Wang_Johnson, Lin} 
However, limited ARPES studies are reported even on Bi$_2$Te$_2$Se to date. \cite{Suyang, Wang_Johnson, Lin, Sergey, BTS_Ando, Ong_BTS, Arakane, Kimura, Souma} In this paper, we investigate the electronic structure of four distinct and unique compounds, namely, Bi$_2$Se$_2$Te (Se-rich), Bi$_2$Te$_2$Se (Te-rich), Bi$_2X_{3-x}$S$_x$ ($X$ = Se, Te; $x$ = 1, 1.5), and GeBi$_2$Te$_4$, as representative members of the ternary family. Surface state properties relevant for the enhanced functionality are identified in these materials. First-principles band calculations are also presented for comparison with our experimental data. 

Our experimental findings are itemized as follows. First, our data suggests that the ternary compound Bi$_2$Se$_2$Te (Se-rich) has a large effective bulk band gap. By tuning the ratio of bismuth to antimony, we are able not only to lower the Fermi level into the band gap but also to fine tune the Fermi level so that it lies exactly at the Dirac point. Second, we show that the Dirac point of Bi$_2$Te$_2$Se (Te-rich) is not isolated from the bulk valence bands when the chemical potential is placed at the Dirac point. Third, we report band structure properties of sulfur doped  Bi$_2X_3$ [Bi$_2X_{3-x}$S$_x$ ($X$ = Se, Te; $x$ = 1, 1.5)] in some detail. The compound Bi$_2$Te$_{1.5}$S$_{1.5}$, derived from Bi$_2$Te$_3$ by replacing Te with S, shows a large bulk band gap and a single Dirac cone surface state, where the Dirac point is located inside the bulk band gap, in contrast to the related Bi$_2$Te$_3$ where the Dirac point is buried inside the bulk valence band. The detail of crystal growth of this compound is described in Ref. [40]. The replacement of Te by S is a critically important process to realize the exposed Dirac point electronic structure in Te-rich sample. Finally, we discuss the electronic structure of GeBi$_2$Te$_4$, which serves as a single Dirac cone topological insulator belonging to a class with nonzero weak topological invariants. Despite its high Te-content, this compound exhibits in-gap Fermi level and isolated Dirac node. This is likely due to the change of global crystal potential associated with the Ge sub-lattice.

\begin{figure}
\includegraphics[width=8.0 cm]{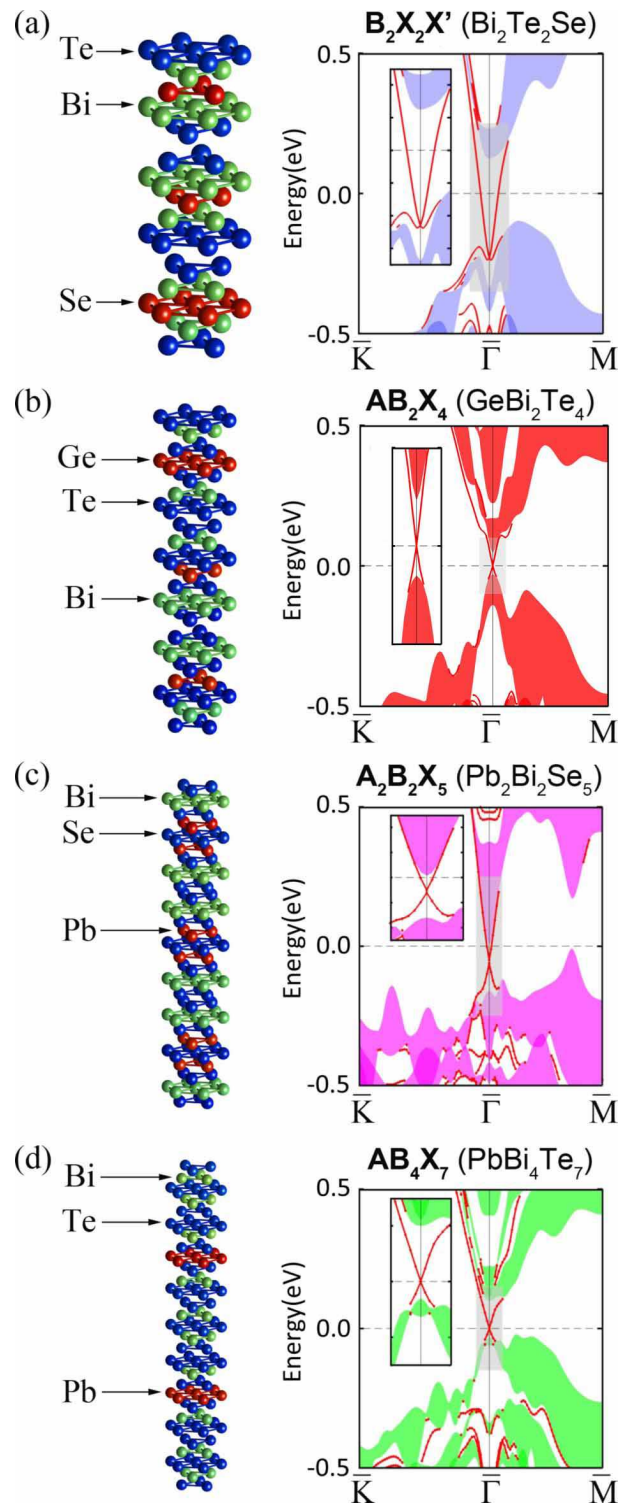}
\caption{(Color online) Crystal structure and topological surface states in ternary spin-orbit compounds: $B_2X_2X'$, $AB_2X_4$, $A_2B_2X_5$ and $AB_4X_7$ [$A$ = Pb, Ge; $B$ = Bi, Sb; $X, X'$ = Se, Te]. (a)-(d) crystal structure and calculated bulk and surface band structures for the (111) surface of $B_2X_2X'$, $AB_2X_4$, $A_2B_2X_5$ and $AB_4X_7$, respectively. The bulk band projections are represented by shaded areas.}
\end{figure}

\section{Methods}


The first-principles band calculations were performed with the linear augmented plane-wave (LAPW) method using the WIEN2K package\cite{wien2k} and the projected augmented wave method\cite{PAW} using the VASP package\cite{VASP} in the framework of density functional theory (DFT). The generalized gradient approximation (GGA) of Perdew, Burke, and Ernzerhof\cite{PBE} was used to describe the exchange correlation potentials. Spin-orbit coupling (SOC) was included as a second variational step using a basis of scalar relativistic eigenfunctions. The surface electronic structure computation was performed with a symmetric slab of six quintuple layers; a vacuum region with thickness larger than 10 $\mathrm{\AA}$ was used.


Single crystalline samples of ternary topological insulators were grown using the Bridgman method, which is described elsewhere. \cite{Hor PRB BiSe, BTS_Ando, Jia} ARPES measurements for the low energy electronic structures were performed at the Synchrotron Radiation Center (SRC), Wisconsin, the Stanford Synchrotron Radiation Lightsource (SSRL), California, and the Advanced Light Source (ALS), California, equipped with high efficiency VG-Scienta SES2002 and R4000 electron analyzers. Samples were cleaved {\it in situ} and measured at 10-80 K in a vacuum better than 1 $\times$ 10$^{-10}$ torr. They were found to be very stable and without degradation for the typical measurement period of 20 hours. Potassium deposition was performed at beam line 12.0.1 of the ALS from a SAES getter source
(SAES Getters USA, Inc.), which was thoroughly degassed before the experiment. Pressure in the experimental chamber stayed below 1$\times$ 10$^{-10}$ torr during deposition. The deposition rate ($\mathrm{\AA}$/Sec) was monitored using commercial quartz thickness monitor (Leybold Inficon Inc., model XTM/2). The deposition amount (thickness) was then obtained by multiplying the deposition rate by the elapsed time.

\begin{figure*}
\centering
\includegraphics[width=16cm]{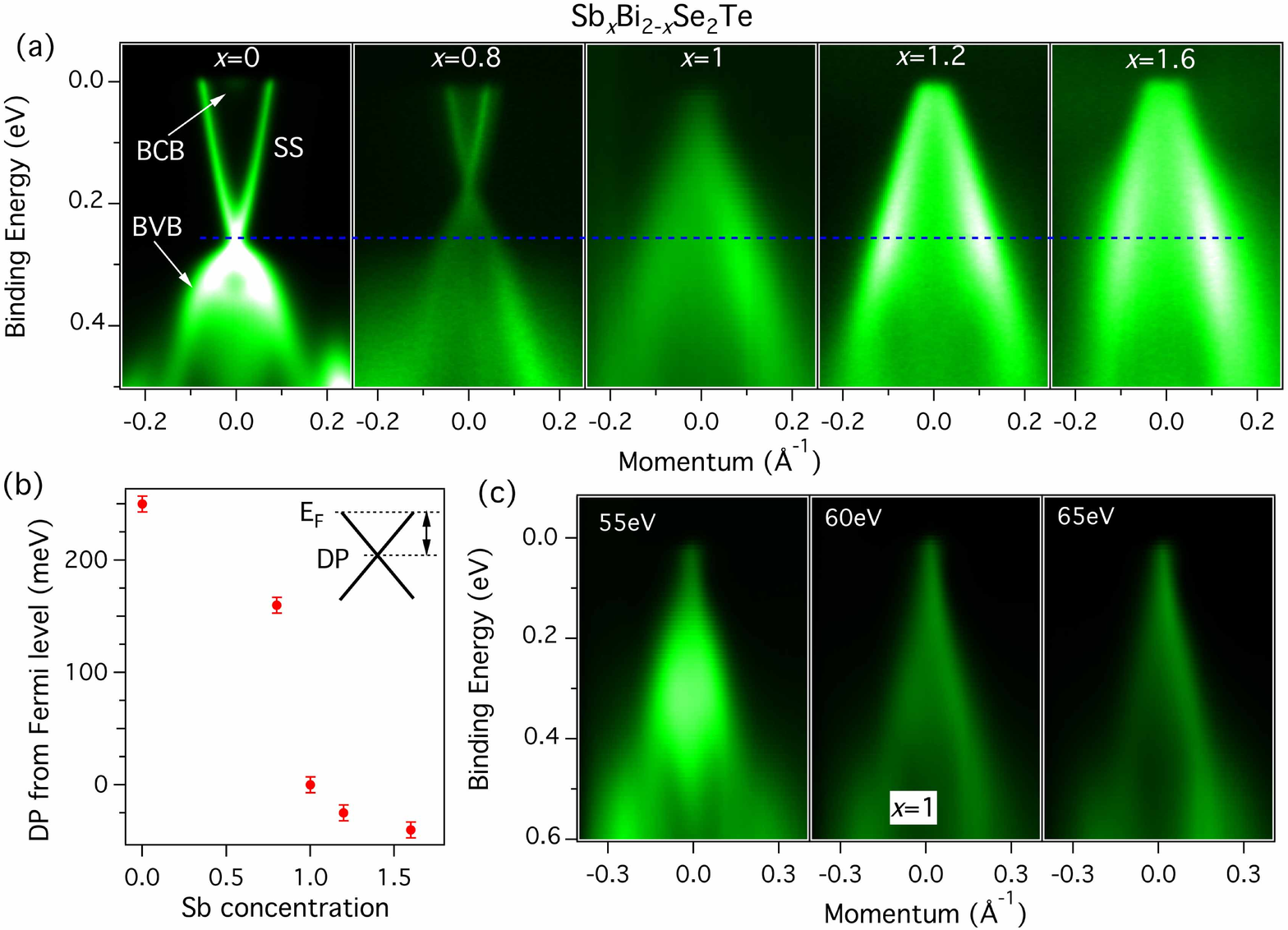}
\caption{Observation of an isolated Dirac node in Sb$_x$Bi$_{2-x}$Se$_2$Te via Dirac point tuning. (a) ARPES $k$-$E$ maps along the $\bar\Gamma$-$\bar M$ momentum direction for Sb$_x$Bi$_{2-x}$Se$_2$Te with $x$ = 0, 0.8, 1, 1.2, and 1.6. The blue dash line marks the binding energy of the Dirac point for $x$ = 0. (b) Binding energy of the Dirac point as a function of $x$. Inset shows the definition of the Dirac point binding energy. (c) Photon energy dependence of ARPES spectra for Sb$_x$Bi$_{2-x}$Se$_2$Te ($x$ = 1). The isolated Dirac point is observed to be in close vicinity to the Fermi level.}
\end{figure*}

\begin{figure*}
\centering
\includegraphics[width=16cm]{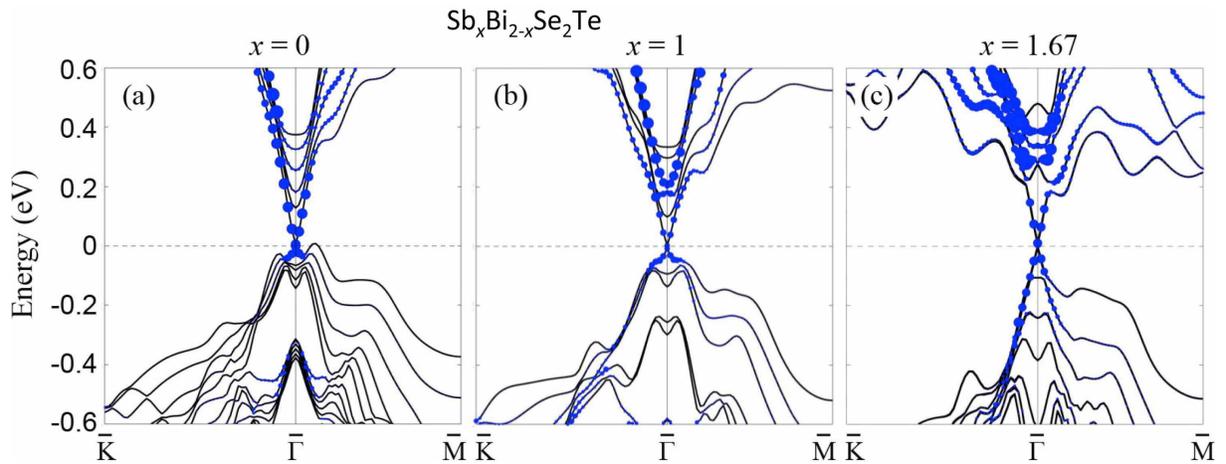}
\caption{First principles band structures of Sb$_x$Bi$_{2-x}$Se$_2$Te based on slab calculations with six quintuple layers for (a) $x$ = 0, (b) $x$ = 1, and (c) $x$ = 1.67. The size of the blue dots represents the fraction of electronic charge residing in the surface layers. All three doping levels show gapless Dirac-cone-like surface bands.}
\end{figure*}

\section{Results and discussion}

\subsection{Band calculation}

In Fig. 1 we present crystal structures and first principles theoretical calculations for the (111) bulk and surface electronic structure of $B_2X_2X'$,  $AB_2X_4$, $A_2B_2X_5$ and $AB_4X_7$ [$A$ = Pb, Ge; B = Bi, Sb; $X, X'$ = Se, Te] as examples for a large family of ternary topological insulators with single Dirac cone. Calculations are presented along the $\bar K-\bar\Gamma-\bar M$ momentum space directions. $B_2X_2X'$ has tetradymite structure with a rhombohedral unit cell belonging to the space group R$\bar3$m. The commonly invoked hexagonal cell consists of three quintuple layers. The natural cleavage plane of Bi$_2$Te$_2$Se lies between two quintuple layers. $A_mB_{2n}X_{m+3n}$ represents a large family of compounds in which $(AX)_m$ layers are inserted into the $B_2X_3$ stacking. The crystal structures of $AB_2X_4$, $A_2B_2X_5$, and $AB_4X_7$ are composed of $X$ layers forming a cubic close packing, with a fraction of octahedral interstices occupied by $A$ and $B$ atoms. \cite{GBT_ Structure} The unit cell of $AB_2X_4$ is formed by stacking together three seven-atomic-layer slabs in the sequence $X(1)-B-X(2)-A-X(2)-B-X(1)$ [Fig. 1(b)]. The cleavage plane of GeBi$_2$Te$_4$ locates between two seven-atomic-layers. Figs. 1(c) and 1(d) give two examples of topologically nontrivial compounds with two different kinds of insertion and stacking. The unit cell of Pb$_2$Bi$_2$Se$_5$ consists of nine-atomic-layers which are made by inserting two PbSe layers into Bi$_2$Se$_3$. This crystal cleaves between two nine-atomic-layers, where the van der Waals bonding is weak. PbBi$_4$Te$_7$ consists of alternating seven-atomic-layers of PbBi$_2$Te$_4$ and quintuple layers of Bi$_2$Te$_3$. There are two possible surface terminations along the (111) direction for PbBi$_4$Te$_7$, with the exposure of either a seven-atomic-layer or a five-atomic-layer. \cite{Suyang, Sergey} We show the surface bands for the one with the exposure of seven-atomic-layer in Fig. 1(d).

A singly degenerate gapless Dirac cone centered at the $\bar{\Gamma}$ point is observed in the representative compounds for each class, indicating that these materials belong to the $Z_2$ = -1 topological class. The numerically predicted bulk band gap varies over an order of magnitude from 0.01 eV to 0.31 eV. Such surface electron kinetics offer a wide range of topologically nontrivial electronic structures ranging from a nearly isotropic Dirac cone (e.g. PbBi$_2$Se$_4$) to strongly anisotropic and doping dependent topological surface states. This remarkable material flexibility provides a wide range of critical electronic properties for realization of different functionalities, which are not even theoretically offered in the most commonly studied Bi$_2X_3$ compounds.

\begin{figure*}
\includegraphics[width=17.5cm]{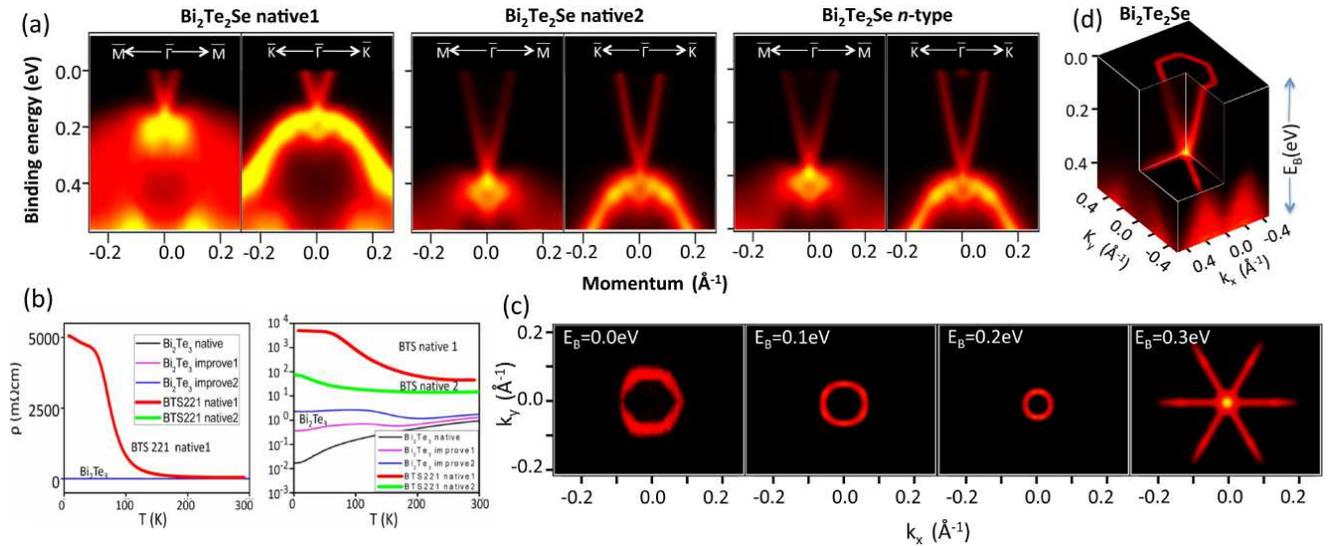}
\caption{(a) ARPES $k$-$E$ maps along the high symmetry directions $\bar{\Gamma}-\bar{M}$ and $\bar{\Gamma}-\bar{K}$ of Bi$_2$Te$_2$Se for three different samples: Bi$_2$Te$_2$Se ``native 1'', Bi$_2$Te$_2$Se ``native 2'', and Bi$_2$Te$_2$Se ``\textit{n}-type'', respectively (see text). Single Dirac cone on the cleaved (111)-surface is observed. 
(b) Resistivity profiles of native Bi$_2$Te$_2$Se, compared to native and improved Bi$_2$Te$_3$, are presented in linear (left) and logarithmic scale (right). 
(c) ARPES maps of constant energy contours presented at several binding energies for the ``native 2'' sample. (d) Three-dimensional illustration of the electronic structure in (c).}
\end{figure*}

\subsection{Realization of an isolated Dirac node}

The presence of an isolated Dirac node, as well as the tunability of the chemical potential to the isolated Dirac point, is highly favored for application purposes because it reduces the scattering from the bulk bands. An important requirement for topological insulators in device oriented applications such as topological quantum information and low power spintronics devices \cite{spintronics} is the dissipationless surface states in the topological transport regime, i.e., an isolated Dirac cone fully separated from bulk bands, and the Fermi level located at the Dirac point.\cite{Liang Fu PRL Superconductivity} The full exposure of topological transport regime for dissipationless spin current with tunable surface states is useful for the study of various novel topological phenomena, such as quantum spin Hall effect, magnetoelectric effects, etc. \cite{Hasan} However, none of the proposed applications have been realized due to the material drawbacks of the existing well-studied topological insulators. Although there are various experimental efforts to realize an isolated Dirac cone by tuning the Fermi level with appropriate doping, \cite{David Nature tunable} the essential necessity of external surface deposition process makes this procedure unsuitable for most practical applications. Recently, tuning of Fermi level has been reported by changing the Bi to Sb composition ratio for Sb$_x$Bi$_{2-x}X_3$ ($X$ = Se, Te) single crystals.\cite{Kong} It is well known that an isolated Dirac node together with a chemical potential lying on the Dirac point through Sb substitution is not possible on either Bi$_{2}$Te$_3$ or Bi$_{2}$Se$_3$. For Sb$_x$Bi$_{2-x}$Te$_3$, though the chemical potential can be tuned by Sb concentration,\cite{Kong} the Dirac point is always buried inside the bulk valence bands. For Sb$_x$Bi$_{2-x}$Se$_3$, substantial Sb substitution changes the topological property of the system since Sb$_2$Se$_3$ is proven to be a trivial insulator.\cite{Hasan, SCZhang}

In the following, we discuss the tunable topological surface states in the Sb$_x$Bi$_{2-x}$Se$_2$Te system, in which we realize an isolated Dirac point without any surface deposition. The ARPES electronic structure of Sb$_x$Bi$_{2-x}$Se$_2$Te is shown in Fig.~2(a). Bi$_{2}$Se$_2$Te ($x$ = 0) shows well-defined surface states with massless Dirac-like dispersion [see Fig. 2(a), left], proving it to be a topological insulator featuring a single Dirac cone with a bulk insulating gap of $\sim$ 250 meV. The Fermi level of this system lies at the bulk conduction band (BCB) and the valence band is located below the Dirac point. Upon substitution of Sb in place of Bi, the Dirac-like topological surface states can be clearly observed in the entire doping range [see Fig.~2(a)]. With increasing $x$, the Fermi level $E_F$ moves downward from the BCB, indicating a reduction of the $n$-type bulk carriers. When Sb substitution is further increased, both Dirac point and $E_F$ lie within the bulk energy gap (such as in $x$ = 0.8), and ultimately the Fermi level reaches the isolated Dirac point for $x$ = 1. Upon further increase of $x$ ($x > 1$), $E_F$ moves below the Dirac point, indicating a crossover from $n$- to $p$-type topological insulator [see Fig.~2(b)]. The charge neutrality point (CNP), the point where $E_F$ meets the Dirac point, can thus be determined to be located at $x \sim 1$. To further verify the observation of an isolated Dirac point, photon energy dependent measurements have been performed, as shown in Fig. 2(c). While bulk valence bands change with photon energy, the surface bands and the isolated Dirac node show no visible dispersion, suggesting the two-dimensional nature of the surface states. Our measurements thus verify that Sb$_x$Bi$_{2-x}$Se$_2$Te is a tunable topological insulator with an isolated Dirac node. Our first-principles calculations for $x$ = 0, 1, 1.67 (see Fig.~3) show that topological surface states exist in all three doping levels; Dirac point moves away from the bulk bands as doping ($x$) increases, supporting our experimental results.

\begin{SCfigure*}
\centering
\includegraphics[width=12.8cm]{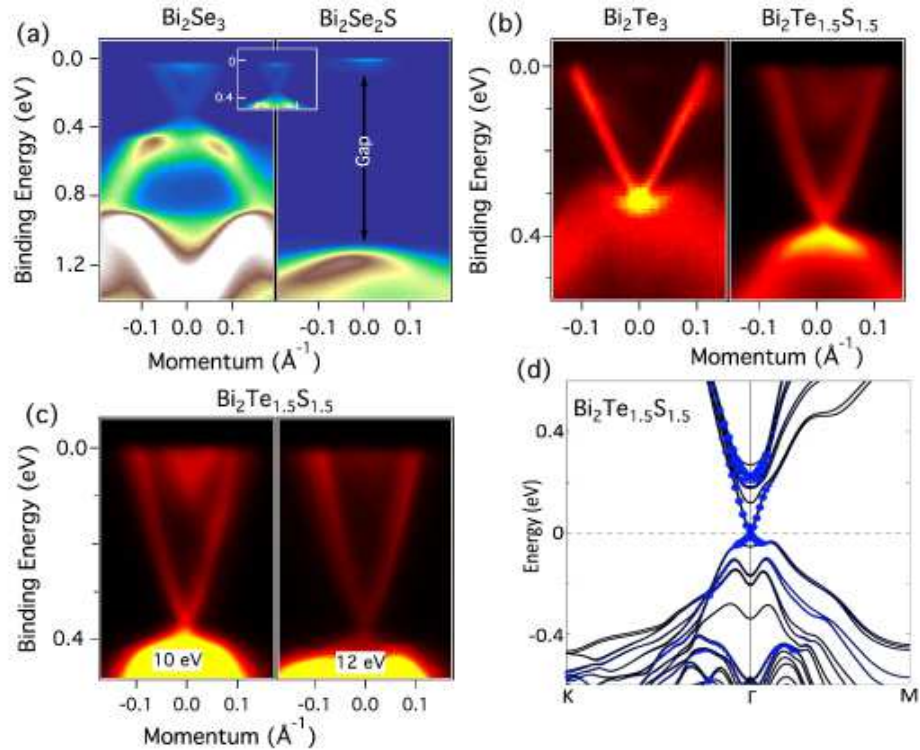}
\caption{Sulfur substitution in Bi$_2X_3$ ($X$ = Se, Te). ARPES $k$-$E$ maps for (a) Bi$_2$Se$_3$ and Bi$_2$Se$_2$S and (b) Bi$_2$Te$_3$ and Bi$_2$Te$_{1.5}$S$_{1.5}$. Inset in (a) shows the zoom of the Bi$_2$Se$_3$ spectrum near the Dirac point. Bi$_2$Se$_2$S is a trivial insulator with a bulk insulating gap of $\sim$ 1.2 eV, while Bi$_2$Te$_{1.5}$S$_{1.5}$ is a topological insulator with single Dirac cone. (c) Photon energy dependence measurements for Bi$_2$Te$_{1.5}$S$_{1.5}$, showing isolated nature of the Dirac point. The measured photon energies are marked on each panel. All data are presented along $\bar\Gamma$-$\bar M$ momentum direction. (d) First-principles band structures of Bi$_2$Te$_{1.5}$S$_{1.5}$ based on slab calculations with six quintuple layers. The size of the blue dots represents the fraction of electronic charge residing in the surface layers.}
\end{SCfigure*}

\subsection{Insulating Bi$_2$Te$_2$Se}

An important property for a functional electronic structure of a topological insulator is the isolation of the surface states from the bulk electronic states, since the surface signal would otherwise be washed out by the bulk contribution in transport experiments. 
Bi$_2$Te$_2$Se is a distinct line compound in the phase diagram known as ``Kawazulite'' \cite{kawazulite, Suyang, BTS_Ando} (as opposed to a random mixture of Bi$_2$Se$_3$ and Bi$_2$Te$_3$). The interpretation of any surface transport measurements will rely on key band structure properties and parameters, such as Fermi velocity, Fermi momentum, etc., which have yet to be reported on Bi$_2$Te$_2$Se. In Fig. 4(a), we present ARPES electronic structures on three batches of as-grown Bi$_2$Te$_2$Se (``native 1'', ``native 2'', and ``$n$-type'') with slightly different growth parameters. Our measurements reveal a single Dirac cone on the cleaved (111)-surface. The experimentally observed chemical potentials vary with different sample growth conditions. As shown in Fig. 4(a), the Fermi level of the ``native1'' batch is slightly above the Dirac node ($E_F=E_D+0.1$ eV) with an average Fermi momentum ($k_F$) of $0.05$ $\mathrm{\AA}^{-1}$. In contrast, the Fermi level of the``native2'' batch is more than 0.3 eV above $E_D$, with a larger averaged $k_F$ of $0.1$ $\mathrm{\AA}^{-1}$. For the batch marked as ``\textit{n}-type'', the bulk conduction band minimum is observed near its Fermi level, from which we are able to obtain a bulk band gap of $\sim0.3$ eV.

The two-dimensional constant energy contour plots of the ARPES intensity at various binding energies ($E_B$) are shown in Fig. 4(c).
The Fermi contour of the ``native2'' batch [first panel of Fig. 4(c)] realizes a hexagonal shape within the bulk band gap. Binding energy evolution study of the constant energy contours [Fig. 4(c)] shows that the hexagon gradually reverts to a circle when approaching the Dirac node. 
In the vicinity of the Dirac point, the valence band feature is observable as a six-fold petal-like intensity pattern at $E_B$ $\sim$ 0.3 eV (Fig. 4(c) right).
Three-dimensional representation of the electronic structure of the ``native2'' batch is shown in Fig. 4(d). From Fig. 4, it is clear that the Dirac point of Bi$_2$Te$_2$Se is not exposed, rather it is buried into the bulk valence bands. The Fermi velocity ($v_F$) of Bi$_2$Te$_2$Se is estimated to be $6 \times 10^5$ m/s along the $\bar{\Gamma}-\bar{M}$ direction, and $8{\times}10^5$ m/s along the $\bar{\Gamma}-\bar{K}$ direction, which is larger than that of Bi$_2X_3$ (Refs. \onlinecite{Matthew Nature physics BiSe,  Chen Science BiTe}). This makes it favorable for a long mean-free- path ($L=v_F\tau$) on the surface.

\subsection{Sulfur to selenium/tellurium substitution in Bi$_2$(Se/Te)$_3$}

The proposed applications of topological insulators require a wide range of tunability of the key electronic parameters of the topological surface states, which is lacking in the widely studied binary TI material Bi$_2$Se$_3$ and Bi$_2$Te$_3$. In the following, we present a study of the sulfur substitution to Se/Te in Bi$_2$Se$_3$ and Bi$_2$Te$_3$ (see Ref. [40] for sample growth and characterization), which brings desired properties to the well-studied binary topological insulators. While Bi$_2$Se$_3$ is a single Dirac cone topological insulator, ARPES measurement shows that Bi$_2$Se$_2$S is a trivial insulator with a band gap of $\sim$ 1.2 eV [see Fig. 5(a)]. On the other hand, ARPES measurement on Bi$_2$Te$_{1.5}$S$_{1.5}$ [Fig~5(b)] reveals that it is a topological insulator with a bulk band gap of $\sim$ 0.2 eV (band gap of Bi$_2$Te$_3$ $\sim$ 0.15 eV). More importantly, compared to Bi$_2$Te$_3$ whose Dirac node is buried inside the valence bands, the Dirac point of Bi$_2$Te$_{1.5}$S$_{1.5}$ is completely isolated from the bulk electronic states, which is essential for applications in the Dirac transport regime. The isolated nature of the Dirac point is further verified by photon energy dependence measurements as shown in Fig. 5(c). The surface states show no visible dispersion, and the Dirac node is always at a different binding energy from that of the bulk valence band under different photon energies. The isolated nature of the Dirac node is also supported by our first principles band calculations presented in Fig. 5(d).

Furthermore, our experimental observations shown in Figs. 5(a) and (b) suggest different roles of sulfur doping on Bi$_2$Se$_3$ and Bi$_2$Te$_3$, which can be understood by considering their crystal structures. While Bi$_2$Se$_3$ has a rhombohedral unit cell under the space group $R\bar{3}m$ (No. 166), Bi$_2$S$_3$ has an orthorhombic unit cell under the space group $Pnma$ (No. 62)(see Ref. \onlinecite{Bi2S3structure}). The former compound is topologically nontrivial, while the latter is a trivial band insulator with a large gap of about 1 eV. The solid solution Bi$_2X_{3-x}$S$_x$ behaves in such a way that formula with higher concentration of heavier elements prefer the rhombohedral structure, while that with higher concentration of lighter elements prefer the orthorhombic structure. The observed large gap in Bi$_2$Se$_2$S is consistent with the predicted trivial phase with an orthorhombic unit cell. On the other hand, Bi$_2$Te$_{1.5}$S$_{1.5}$ contains heavier elements and our observed gapless surface Dirac cone is consistent with the predicted nontrivial phase with a rhombohedral unit cell.

\subsection{GeBi$_2$Te$_4$: A ternary topological insulator with nonzero weak topological invariants}

\begin{figure*}
\centering
\includegraphics[width=15cm]{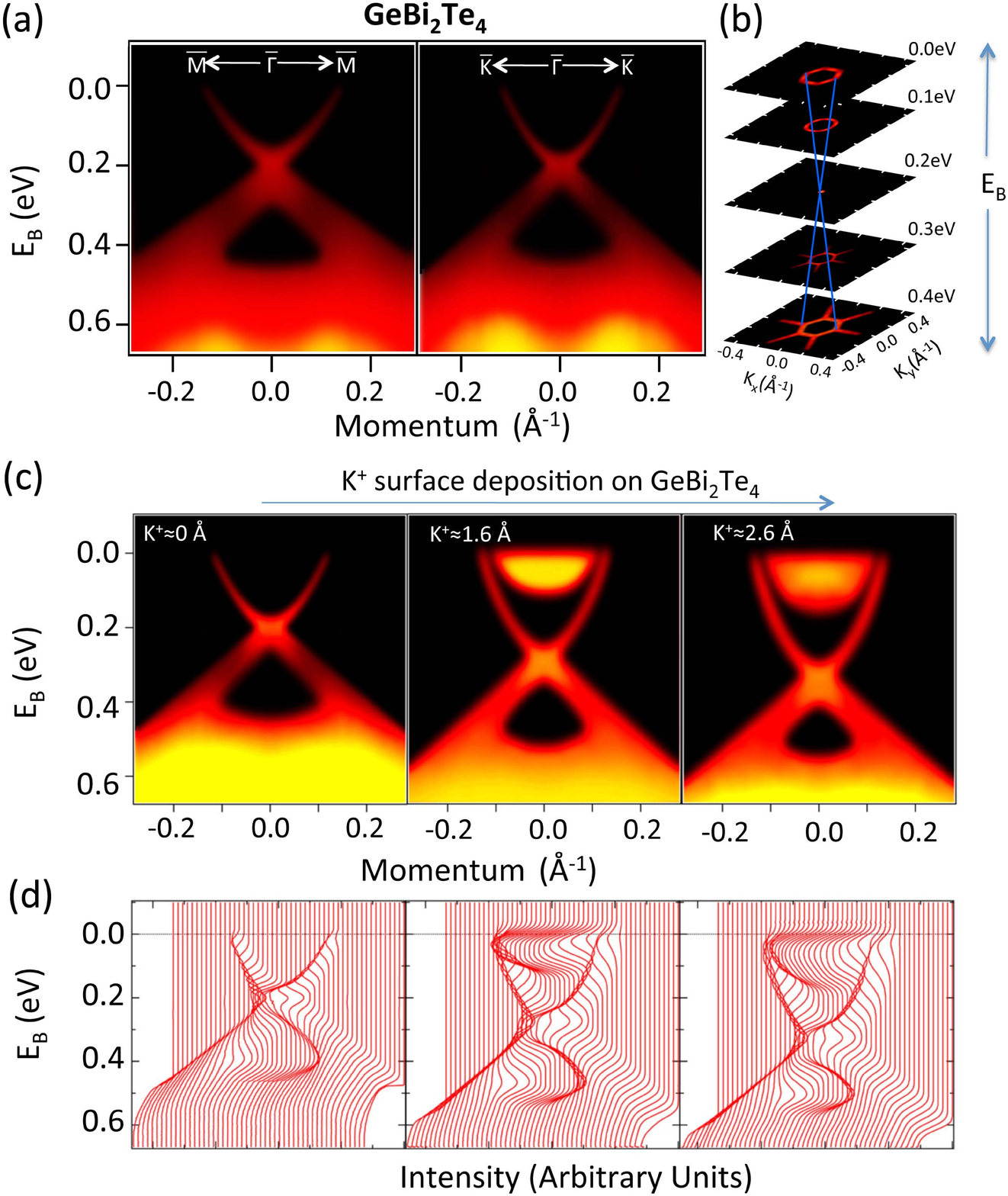}
\caption{(Color online) (a) ARPES $k$-$E$ maps of GeBi$_2$Te$_4$ along $\bar M$-$\bar\Gamma$-$\bar M$ and $\bar K$-$\bar\Gamma$-$\bar K$ high symmetry directions. (b) ARPES intensity maps for various binding energies between $E_F$ and 0.4 eV. (c) ARPES spectra of GeBi$_2$Te$_4$ along the $\bar\Gamma$-$\bar K$ direction as potassium is deposited. Average thickness of the deposition layer is indicated on the top of each panel. (d) Energy distribution curves (EDCs) for the ARPES spectra shown in panel (c).}
\end{figure*}

Topological insulators are characterized by $Z_2$ topological invariants of the bulk band structure. For a three-dimensional bulk insulator, the topological insulator state is defined by four topological invariants commonly indexed as $[\nu_0;\nu_1\nu_2\nu_3]$ (see Refs. \onlinecite{Kane PRL, David Nature08}), where $\nu_0$ is the strong invariant and $ \nu_1,\nu_2,\nu_3$ are the weak invariants. In the bulk computations, we evaluate these four invariants by obtaining the parity symmetry of the Bloch wave functions for all the occupied electronic states at eight time-reversal-invariant points.\cite{FuKane} While Bi$_2$Se$_3$ belongs to the [1;000] class due to a band inversion at the $\Gamma$ point, GeBi$_2$Te$_4$ is predicted to be a strong topological insulator with nonzero weak indices [1;111]. This is due to a band inversion at the $Z$-point instead of the $\Gamma$-point, \cite{FuKane} which is also seen in PbBi$_2$Te$_4$.

Figure 6(a) shows the ARPES measured dispersion of the surface Dirac bands of GeBi$_2$Te$_4$. Our data shows an in-gap Fermi level for naturally grown GeBi$_2$Te$_4$ crystals. In order to systematically analyze the surface band structure and Fermi surface warping effects in GeBi$_2$Te$_4$, we plotted constant energy contours at different binding energies [see Fig. 6(b)]. The constant energy contour at $E_B$ = 0.02 eV of GeBi$_2$Te$_4$ clearly demonstrates the hexagonal warping effect. \cite{Liang_Fu, Xu_BiTe} When the binding energy is increased from the Fermi level, the effect of the bulk potential vanishes and the shape of the contour turns into a circle. Further increasing the binding energy results in a Fermi surface consisting of a single Dirac point with no other features. Therefore, GeBi$_2$Te$_4$ realizes an isolated Dirac point, which makes it possible to bring the system into the topological transport regime. Constant energy contours below the Dirac point reveal the lower Dirac cone. Upon further increasing the binding energy, an additional six-fold symmetric feature extending outwards along all $\bar\Gamma$-$\bar M$ directions is observed.

Surface potassium deposition measurements are performed in order to estimate the energy position of the bottom of the bulk conduction band from the Dirac point. Fig. 6(c) shows the ARPES measured dispersion of the surface bands of GeBi$_2$Te$_4$ along the $\bar\Gamma$-$\bar K$ momentum direction as potassium is deposited; corresponding energy distribution curves are plotted in Fig. 6(d). The average thickness of the potassium layer is also marked on the top of each panel in Fig. 6(c). With approximately 1.6 $\mathrm{\AA}$ deposition of potassium, the bulk conduction band appears with its bottom located at about 200 meV above the Dirac point.

\section{Conclusion}

We have performed electronic structure measurements for representative members of a large family of ternary topological insulators using ARPES.
Our measurements show that the ternary topological insulators Bi$_2$Se$_2$Te, Bi$_2$Te$_2$Se, Bi$_2$Te$_{1.5}$S$_{1.5}$ and GeBi$_2$Te$_4$ exhibit single Dirac cone surface states, which is supported by our first principles band calculations. Among them, Bi$_2$Se$_2$Te, Bi$_2$Te$_{1.5}$S$_{1.5}$ and GeBi$_2$Te$_4$ feature an in-gap Dirac point. Bi$_2$Se$_2$Te has a large effective bulk band gap and it shows a tunable surface state with an isolated Dirac node upon changing chemical composition Bi/Sb. The unique electronic properties of this material class identified in our experiments will be a helpful guide to interpret transport, optical, magnetic and thermoelectric measurements.

\section{Acknowledgements}

The ARPES measurements are supported by NSF-DMR-1006492. The Synchrotron Radiation Center is supported by NSF-DMR-0537588. The crystal growth is supported by NSF-DMR-0819860. Work at Northeastern is supported by the Basic Energy Sciences, US Department of Energy (DE-FG02-07ER46352 and AC03-76SF00098), and benefited from the allocation of supercomputer time at NERSC and Northeastern University's Advanced Scientific Computation Center. T.-R.C. and H.T.J. are supported by the National Science Council and Academia Sinica, Taiwan, and they thank NCHC, CINC-NTU, and NCTS, Taiwan for technical support. The Advanced Light Source is supported by the Director, Office of Science, Office of Basic Energy Sciences, of the U.S. Department of Energy under Contract No. DE-AC02-05CH11231. The Stanford Synchrotron Radiation Lightsource is supported by the U.S. Department of Energy under Contract No. DE-AC02-76SF00515. M.Z.H. acknowledges additional support from the Advanced Light Source at LBNL and the A. P. Sloan Foundation.


\begin{thebibliography}{99}

\bibitem{Kane PRL} L. Fu, C. L. Kane, and E. J. Mele, Phys. Rev. Lett. $\mathbf{98}$, 106803 (2007).

\bibitem{David Nature08} D. Hsieh, D. Qian, L. Wray, Y. Xia, Y. S. Hor, R. J. Cava, and M. Z. Hasan, Nature $\mathbf{452}$, 970 (2008).

\bibitem{Suyang_1} S.-Y. Xu, L. A. Wray, Y. Xia, R. Shankar, A. Petersen, A. Fedorov, H. Lin, A. Bansil, Y. S. Hor, D. Grauer, R. J. Cava, and M. Z. Hasan, arXiv:cond-mat/1007.5111v1 (2010). e-print arXiv:1007.5111v1


\bibitem{Ran Nature physics} Y. Ran, Y. Zhang, and A. Vishwanath, Nature Phys. $\mathbf{5}$, 298 (2009).

\bibitem{David Science BiSb} D. Hsieh, Y. Xia, L. Wray, D. Qian, A. Pal, J. H. Dil, J. Osterwalder, F. Meier, G. Bihlmayer, C. L. Kane, Y.S. Hor, R. J. Cava, and M. Z. Hasan, Science $\mathbf{323}$, 919 (2009).

\bibitem{Hasan} M. Z.  Hasan and C. L. Kane, Rev. Mod. Phys. $\mathbf{82}$, 3045 (2010).

\bibitem{SCZhang}X.-L. Qi, and S. C. Zhang, Rev. Mod. Phys. $\mathbf{83}$, 1057 (2011).

\bibitem{Matthew Nature physics BiSe} Y. Xia, D. Qian, D. Hsieh, L. Wray, A. Pal, H. Lin, A. Bansil, D. Grauer, Y. S. Hor, R. J. Cava, and M. Z. Hasan, Nature Phys. $\mathbf{5}$, 398 (2009).

\bibitem{Chen Science BiTe} Y. L. Chen, J. G. Analytis, J.-H. Chu, Z. K. Liu, S.-K. Mo, X. L. Qi, H. J. Zhang, D. H. Lu, X. Dai, Z. Fang, S. C. Zhang, I. R. Fisher, Z. Hussain, and Z.-X. Shen, Science $\mathbf{325}$, 178 (2009).

\bibitem{David Nature tunable} D. Hsieh, Y. Xia, D. Qian, L. Wray, J. H. Dil, F. Meier, J. Osterwalder, L. Patthey, J. G. Checkelsky, N. P. Ong, A. V. Fedorov, H. Lin, A. Bansil, D. Grauer, Y. S. Hor, R. J. Cava, and M. Z. Hasan, Nature $\mathbf{460}$, 27 (2009).

\bibitem{Pedram Nature BiSb} P. Roushan, J. Seo, C. V. Parker, Y. S. Hor, D. Hsieh, D. Qian, A. Richardella, M.Z. Hasan, R. J. Cava, and A. Yazdani, Nature $\mathbf{460}$, 1106 (2009).

\bibitem{Hor PRB BiSe} Y. S. Hor, A. Richardella, P. Roushan, Y. Xia, J. G. Checkelsky, A. Yazdani, M. Z. Hasan, N. P. Ong, and R. J. Cava, Phys. Rev. B $\mathbf{79}$, 195208 (2009).

\bibitem{Essin PRL Magnetic} A. M. Essin, J. E. Moore, and D. Vanderbilt, Phys. Rev. Lett. $\mathbf{102}$, 146805 (2009).

\bibitem{Galvanic effect} I. Garate and M. Franz, Phys. Rev. Lett. $\mathbf{104}$, 146802 (2010).

\bibitem{Yu Science QAH} R. Yu, W. Zhang, H.-J. Zhang, S.-C. Zhang, X. Dai, and Z. Fang, Science $\mathbf{329}$, 61 (2010).

\bibitem{Qi Science Monopole} X.-L. Qi, R. Li, J. Zhang, and S.-C. Zhang, Science $\mathbf{323}$, 1184 (2009).

\bibitem{Linder PRL Superconductivity} J. Linder, Y. Tanaka, T. Yokoyama, A. Sudbo, and N. Nagaosa, Phys. Rev. Lett. $\mathbf{104}$, 067001 (2010).

\bibitem{Liang Fu PRL Superconductivity} L. Fu and C. L. Kane, Phys. Rev. Lett. $\mathbf{102}$, 216403 (2009).

\bibitem{Phuan} D.-X. Qu, Y. S. Hor, J. Xiong, R. J. Cava, and N. P. Ong, Science $\mathbf{329}$, 821 (2010).

\bibitem{Hor arXiv BiTe superconducting} Y. S. Hor, J. G. Checkelsky, D. Qu, N. P. Ong, and R. J. Cava, arXiv:1006.0317 (2010).

\bibitem{Moore} P. Ghaemi, R. S. K. Mong, and J. E. Moore, arXiv:cond-mat/1002.1341v1 (2010).

\bibitem{Suyang} S.-Y. Xu, L. A. Wray, Y. Xia, R. Shankar, A. Petersen, A. Fedorov, H. Lin, A. Bansil, Y. S. Hor, D. Grauer, R. J. Cava, and M. Z. Hasan, arXiv:cond-mat/1007.5111v1 (2010).
 
\bibitem{Wang_Johnson} L.-L. Wang, and D. D. Johnson, Phys. Rev. B, $\mathbf{83}$, 241309(R) (2011).

\bibitem{Lin} H. Lin, T. Das, L. A. Wray, S.-Y. Xu, M. Z. Hasan, and A. Bansil, New J. Phys. $\mathbf{13}$, 095005 (2011).

\bibitem{BTS_Ando} Z. Ren, A. A. Taskin, S. Sasaki, K. Segawa, and Y. Ando, Phys. Rev. B $\mathbf{82}$, 241306(R) (2010).

\bibitem{Ong_BTS} J. Xiong, Y. Luo, Y. Khoo, S. Jia, R. J. Cava, and N. P. Ong, arXiv:1111.6031 (2011).

\bibitem{Sergey} S. V. Eremeev, G. Landolt, T. V. Menshchikova, B. Slomski, Y. M. Koroteev,
Z. S. Aliev, M. B. Babanly, J. Henk, A. Ernst, L. Patthey, A. Eich,
A. A. Khajetoorians, J. Hagemeister, O. Pietzsch, J. Wiebe, R. Wiesendanger,
P. M. Echenique, S. S. Tsirkin, I. R. Amiraslanov, J. H. Dil, and E. V. Chulkov, Nature com., $\mathbf{3}$, 635 (2012).

\bibitem{Arakane} T. Arakane, T. Sato, S. Souma, K. Kosaka, K. Nakayama, M. Komatsu, T. Takahashi, Z. Ren, K. Segawa, and Y. Ando, Nature com., $\mathbf{3}$, 636 (2012).

\bibitem{Kimura} K. Kuroda, H. Miyahara, M. Ye, S. V. Eremeev, Yu. M. Koroteev, E. E. Krasovskii, E. V. Chulkov, S. Hiramoto, C. Moriyoshi, Y. Kuroiwa, K. Miyamoto, T. Okuda, M. Arita, K. Shimada, H. Namatame, M. Taniguchi, Y. Ueda, and A. Kimura, {Phys. Rev. Lett.} \textbf{108}, 206803 (2012).

\bibitem{Souma} S. Souma, K. Eto, M. Nomura, K. Nakayama, T. Sato, T. Takahashi, K. Segawa, and Y. Ando, Phys. Rev. Lett., \textbf{106}, 216803 (2011).

\bibitem{wien2k}P. Blaha, K. Schwarz, G. Madsen, D. Kvasnicka, and J. Luitz, \textit{WIEN2k, An Augmented Plane Wave Plus Local Orbitals Program for
Calculating Crystal Properties.} (Karlheinz Schwarz, Techn. University Wien, Austria, 2001).

\bibitem{PAW}
P. E. Bl$\ddot{o}$chl, Phys. Rev. B. {\bf 50}, 17953 (1994); G. Kresse and J. Joubert, Phys. Rev. B. {\bf 59}, 1758 (1999).

\bibitem{VASP}
G. Kress and J. Hafner, Phys. Rev. B. {\bf 48}, 13115 (1993); G. Kress and J. Furthm$\ddot{u}$ller, Comput. Mater. Sci. {\bf 6}, 15 (1996); Phys. Rev. B. {\bf 54}, 11169 (1996).

\bibitem{PBE} J. P. Perdew, K. Burke, and M. Ernzerhof, {Phys. Rev. Lett.} \textbf{77}, 3865-3868 (1996).

\bibitem{Jia} S. Jia, H. Ji, E. Climent-Pascual, M. K. Fuccillo, M. E. Charles, J. Xiong, N. P. Ong, and R. J. Cava, Phys. Rev. B $\mathbf{84}$, 235206 (2011).

\bibitem{GBT_ Structure} K. A. Agaev and S. A. Semiletov, Kristallografiya $\mathbf{10}$, 109 (1965).

\bibitem{spintronics} S. A. Wolf, D. D. Awschalom, R. A. Buhrman, J. M. Daughton, S. von Molnar, M. L. Roukes, A. Y. Chtchelkanova, and D. M. Treger, Science $\mathbf{294}$ 1488 (2001).

\bibitem{Kong} D. Kong, Y. Chen, J. J. Cha, Q. Zhang, J. G. Analytis, K. Lai, Z. Liu, S. S. Hong, K. J. Koski, S.-K. Mo, Z. Hussain, I. R. Fisher, Z.-X. Shen, and Y. Cui, Nature nanotech, $\mathbf{6}$, 705 (2011).

\bibitem{kawazulite} P. Bayliss, American Mineralogist $\mathbf{76}$, 257 (1991).

\bibitem{Huiwen} H. Ji, J. M. Allred,  M. K. Fuccillo,  M. E. Charles, M. Neupane,  L. A. Wray,  M. Z. Hasan,  and R. J. Cava, Rev. B $\mathbf{85}$, 201103(R) (2012).

\bibitem{Bi2S3structure} L. F. Lundegaard, E. Makovicky, T. Boffa-Ballaran, T. Balic-Zunic, Phys. Chem. Miner. $\mathbf{32}$, 578 (2005).

\bibitem{FuKane} L. Fu, and C. L. Kane, Phys. Rev. B, \textbf{76}, 045302 (2007).


\bibitem{Liang_Fu} L. Fu, Phys. Rev. Lett., \textbf{103}, 266801 (2009).


\bibitem{Xu_BiTe} S.-Y. Xu, L. A. Wray, Y. Xia, F. von Rohr, Y. S. Hor, J. H. Dil, F. Meier, B. Slomski, J. Osterwalder, M. Neupane, H. Lin, A. Bansil, A. Fedorov, R. J. Cava, and M. Z. Hasan, arXiv:1101.3985 (2011) (unpublished).

\end{thebibliography}
\end{document}